\documentclass[sigconf,manuscript]{acmart}
\usepackage{natbib}
\AtBeginDocument{%
  }

\setcopyright{acmlicensed}
\copyrightyear{2026}
\acmYear{2026}
\acmDOI{XXXXXXX.XXXXXXX}
\acmConference[CHI'26]{CHI2026 Workshops - From Papers to the Real World: Making Fabrication Research Matter}{April 12--17,
  2026}{Barcelona, Spain}
\begin{document}

\title{The Renaissance of Repair: A Timely Opportunity for Fabrication Research }


\author{Julian Britten}
\email{julian.britten@uni-ulm.de}
\orcid{0000-0002-2646-2727}
\affiliation{%
  \institution{Institute of Media Informatics, Ulm University}
  \city{Ulm}
  \country{Germany}
}

\author{Jan Henry Belz}
\orcid{0000-0003-4628-6107}
\affiliation{%
  \institution{Dr. Ing. h.c. F. Porsche AG}
  \city{Stuttgart}
  \country{Germany}
}
\email{jan_henry.belz@porsche.de}

\renewcommand{\shortauthors}{Britten and Belz}

\begin{abstract}
Through the rise of the right-to-repair movement, along with supporting legislation, we are currently witnessing an attitude shift in favor of repairing. This opens up various opportunities for personal fabrication research. Although the field has shifted more towards sustainable practices, repair is rarely the main focus. In this paper, we want to make the case for repair-centered fabrication research as a timely, relevant, impactful, and therefore meaningful topic. We describe potential avenues researchers could pursue by defining repair as a five-step process, including issue identification, exploring solutions, acquiring materials, performing the repair, and testing, and discuss challenges and opportunities for each step.
\end{abstract}





\maketitle

\section{Introduction}
%wip
%repairing was made more difficult on purpose (break on purpose, warranty voiding, difficult assembly, no information on replacement parts)
%repair is important
%repair is sustainable
%repair is so back 

%- describe right to repair movement, 

The act of repairing has become a task that has fallen out of fashion over the last few decades. Products have increased in complexity significantly, which makes repairs more complex, as they require specialized tools and expertise. Various practices exist to discourage consumers from repair: Sometimes, products come with intended breaking points or are designed with planned obsolescence~\cite{bisschop_designed_2022, lu_unmaking_2024} in mind. Spare parts often don't exist or aren't made available~\cite{schulze_obsolescence_2012}. Sometimes, complexity is artificially increased: In 2026, BMW patented a new screw design representing their logo\footnote{\url{https://www.carscoops.com/2025/12/bmw-just-designed-a-screw-that-locks-you-out-of-your-own-repairs/}, accessed: 10.02.2026} which requires specialized tools. If used in future vehicles, it may prevent owners from doing their own maintenance or repairs. On occasion, consumers are scared into buying new instead of repairing: Tactics range from claiming user-initiated repairs are dangerous~\cite{noauthor_apple_2021} to using illegal "warranty-void" stickers~\cite{aragon_warranty_2026}. Ultimately, all this drives up the price for a repair service nearing the cost crossover point, where consumers would rather buy new than repair. The ecological footprint caused by this is noticeable: On average, household electronics like TVs or vacuum cleaners are used 2.3 years shorter than their designed or desired lifetimes~\cite{noauthor_europes_2020}. 

However, nowadays, consumers are more aware of sustainable practices, and the right to repair movement is growing: From farmers fighting for their right to repair their own tractors~\cite{noauthor_us_2023} to companies like Apple beginning to offer self-service repair kits~\cite{noauthor_apple_2021}, the "battle over the right to repair is tipping in favor of consumers"~\cite{kugler_fight_2023}. In 2020, a study~\cite{noauthor_special_2020} found that 78\% of EU citizens think manufacturers should be required to make it easier to repair digital devices. However, this number drops to only 24\% if it leads to increased prices.
Legislators are responding to this attitude shift: In 2024, the European Parliament voted in favor of a "right to repair" directive~\cite{noauthor_right_2024} which members must transpose into national law by July 2026. 
Similar to Minnesota's New Digital Fair Repair Act \footnote{\url{https://www.ag.state.mn.us/Consumer/Publications/RightToRepair.asp}, accessed 09.02.2026}, it is designed to promote the repair of products. It requires manufacturers to offer repairs within a reasonable timeframe and provide access to spare parts, tools, and repair information. Further, it requires companies to allow the use of 3D-printed or second-hand replacement parts. 
Similar to how previous EU directives, such as the common phone chargers directive \footnote{\url{https://commission.europa.eu/news-and-media/news/eu-common-charger-rules-power-all-your-devices-single-charger-2024-12-28_en}, accessed: 09.02.2026} have pressured the global market into adopting these standards worldwide, we believe that this directive will affect product design and promote repair globally. %i think based on context this is enough
%- eu law: good first start, but just because repairability is required for new products, we wont throw away all of our old stuff

While this political and societal shift is relatively recent, working with existing materials and salvaged components has long been common practice in DIY and maker communities. Makers frequently repurpose old parts, reuse spare components, or fabricate replacements themselves - not only as an expression of creativity and autonomy, but also as a cost-saving measure~\cite{vyas_democratizing_2023}. 
In this sense, repair has always been a part of DIY and, as such, tied to personal fabrication from the beginning~\cite{mota_rise_2011}, and can be promoted further through "new" techniques: 3D printers are great for recreating a small plastic piece that went missing ages ago or fixing a very specific issue through a custom part~\cite{van_oudheusden_3d_2023}. 

However, repair through personal fabrication is by no means a "solved issue". As identified by \citet{yan_make_2025}, there is a lack of repair and maintenance tools. Yet, through the DIY community, the internet is filled with resources:
Online repositories such as \href{https://makerworld.com/en/search/models?keyword=replacement+part&orderBy=downloadCount}{MakerWorld}, \href{https://www.thingiverse.com/search?q=replacement+part&page=1&sort=relevant}{Thingiverse}, or \href{https://www.printables.com/model?category=51}{Printables}, are filled with replacement parts designed by others. \href{https://www.ifixit.com/}{iFixit} and \href{https://www.instructables.com/search/?q=repair&projects=featured}{Instructables} offer thousands of repair manuals and guides for free. 
YouTube is filled with creators like \href{https://www.youtube.com/@JoeyDoesTech}{JoeyDoesTech} showing how easy and fast electronics repair can be. While they are available, it is currently up to the user to seek out these resources. Researchers could therefore develop workflows that simplify the repair process for personal fabrication for novices, experts, and makerspaces.

Further, while the right to repair EU directive influences how new products released after June 2025 need to be designed, old products remain unaffected~\cite{noauthor_right_2024}. Yet, they will stay in circulation for the foreseeable future and require repairs as well. 
Thus, two avenues can be supported: Assisting in the creation of new, more repairable products~\cite{kilic_user-centred_2024} and developing workflows that assist users in both new and old devices. Ultimately, both lead towards more sustainable product lifecycles. 
% rephrased to sound less like strawman ^
%could make an enemy for life and lean into baudisch's "personal fabrication" book mentioning "repair" once. also misses to connect it to sustainability which it highlights as a challenge. maybe there is a nice way to say this? :>
%In their work "Personal Fabrication", \citet{baudisch_personal_2017} identify "sustainability" as . 
%fabrication research hype topic: sustainable rapid prototyping
%We can clearly observe a shift in fabrication research in recent years, as more and more publications focus specifically on introducing more sustainable practices, e.g., through sustainable rapid prototyping~\cite{yan_dissolvpcb_2025}, enabling recycling~\cite{wen_enabling_2025}, reusing existing~\cite{lu_protopcb_2025, mei_fabhacks_2024, iyer_xr-penter_2025} or utilizing alternative materials~\cite{perroni-scharf_sustainaprint_2025, bell_3d_2025}. 
%fabrication research doesnt really focus on repair, yet the practice is as sustainable as it gets
%often as a sidenote or side use case, but rarely in the focus

Yet, even though there is an observable shift towards sustainability-focused research - a key challenge for long-term success for the field~\cite{baudisch_personal_2017}-, e.g., through sustainable rapid prototyping~\cite{yan_dissolvpcb_2025}, enabling recycling~\cite{wen_enabling_2025}, reusing existing~\cite{lu_protopcb_2025, mei_fabhacks_2024, iyer_xr-penter_2025} or utilizing alternative materials~\cite{perroni-scharf_sustainaprint_2025, bell_3d_2025}, fabrication research rarely specifically focuses on repair and often only mentions it as a secondary or tertiary use case. %it would be awesome to take the last X years of CHI and look how many papers focus specifically on repair, but maybe we can get away with it anyways
%-> explore why? not "sexy" enough?
%-> 3d printing sometimes seen as "just creating plastic waste quickly" -> for repair it would outweigh the ecological footprint of buying a new thing basically every time 
We find this somewhat surprising, as repairing might be the most sustainable use case for personal fabrication overall: Instead of ordering a replacement part from the manufacturer online and shipping it around the world, consumers can print the part at home or at a nearby repair shop or maker space. Instead of shipping a broken smartphone to a remote location, it can be repaired locally. This claim is supported by supply chain research: For example,~\citet{li_additive_2017} found that additive-manufacturing-based supply chains reduce both variable costs and carbon emissions relative to conventional replacement part logistics, quantitatively demonstrating sustainability advantages for distributed fabrication approaches in spare-parts provisioning.

However, repair is a multifaceted process that goes beyond the actual repair task itself and can be split into multiple steps (issue identification, exploring solutions, acquiring materials, perform the repair, and testing), each of which provides opportunities for fabrication research, which we outline in this paper.
We believe that to "make fabrication research matter", the community should focus on meaningful use cases with real-world applicability. The current "renaissance of repair" driven by the right to repair movement and supporting legislation that bets on a resurgence/increase of repair shops and repair cafes~\cite{noauthor_right_2024}, provides the perfect opportunity for fabrication researchers: We can help shape the repair process for future devices, but also support the upkeep and maintenance of older objects where new legislation, such as the EU "right to repair" directive, does not apply.

%idea: tie it how we were removed further and further from the fabrication process of goods and dont know how what works and how much work actually went into making something (alienation) %fuck, this is marxism again
%so that today the average person doesnt know -> maybe we can change that. (Jasmine Lu "recomputing e-waste" as example)

\section{The Repair Process - Challenges and Opportunities}

\begin{figure}[H]
  \includegraphics[width=\textwidth]{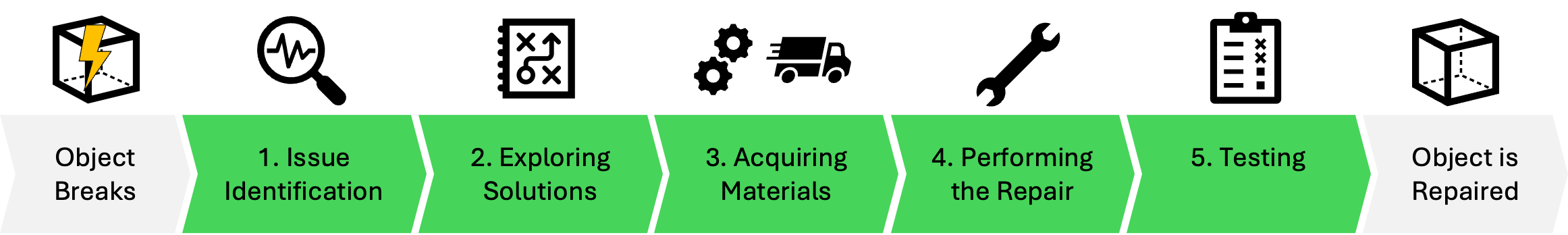}
  \caption{The repair task defined as a five-step process.}
  \Description{A flow chart visualizing the five-step process. After an object breaks, we first have to identify the issue, then explore possible solutions. The third step includes acquiring materials in various ways, and in the fourth step, the actual repair is performed. Finally, the object is tested before we define it as repaired.}
  \label{fig:process}
\end{figure}

The repair process goes beyond the actual repair task (e.g., replacing a broken piece). We define it as a five-step process (see \autoref{fig:process}), each of which offers opportunities for fabrication research. In this section, we want to outline said steps and discuss opportunities and challenges.

%could make it five if we include test
%but repair comes with multi-faceted challenges (Fehlererkennung, Lösungsvorschläge bringen, Teile herstellen, Reparatur durchführen, testen)
\subsection{Issue Identification}
First, the issue with a broken or malfunctioning object has to be identified. A key challenge here is to help users navigate the wide variety of devices or objects of varying complexity, along with different versions with minor differences, and guide them through the troubleshooting steps. A recent work by \citet{kwatra_splatoverflow_2025} attempts this by overlaying a Gaussian splat of a faulty device with a "correct" 3D model. Through the assistance of an expert, the user is guided through troubleshooting steps. Future work could iterate on this, e.g., by eliminating the need for a human expert.

\subsection{Exploring Solutions}
%- finding solution for "old" difficult to repair tech
%- beating Amazon prime (fixing has to be faster than buying new), convincing users to act sustainable
%we envision recommender system (technically a lie, we dont do it rn but i would want to)
% find optimal use case
Next, the user must be presented with the optimal step moving forward. Challenges in this step include finding solutions for older objects (e.g., availability of spare parts or documentation), but also convincing users that repair is worth the effort and feasible. Beyond the moral sustainability argument, we have to consider user effort, skill level, time, available tools, and cost~\cite{rosner_designing_2014}. We envision a system that recommends multiple pathways, possibly utilizing nudging strategies~\cite{valta_digital_2025}.

%designing part, physically creating part (working within 3d printing constraints), assembly instructions
%- BA Lucas (abpausen)
%- BA Nils (how to adapt parts for printing process)
\subsection{Acquiring Materials}
The third step in the repair process involves acquiring the necessary parts and materials for the repair task, such as designing and producing replacement parts using personal fabrication devices. Previous research on alternative design processes that minimize required expertise skills (e.g., parametric design tools~\cite{stemasov_param_2024}) could be revisited for applicability in the repair context. Additionally, alternative sustainable filaments like eggshells~\cite{bell_3d_2025} could be explored specifically with a focus on repairability. A major challenge for this step is constraints of personal fabrication techniques like additive manufacturing: 3D-printed parts may need to be adjusted compared to the original, as different physical parameters (layer height, print speed, sagging, etc.) need to be considered. Non-experts may not be aware of this and would need supporting tools to avoid frustration. Ongoing works in our lab explore these aspects.

\subsection{Performing the Repair}
%- BA Benedict (find how to do instructions for repair)
In the fourth step, the actual repair is taking place. Here, we can explore how users can be guided through the repair process while considering their skill level and expertise. First-time repairers may require more assistance and detailed steps, whereas experts may prefer more abstract, to-the-point instructions. Related research has explored how augmented reality (AR) can be used for task guidance~\cite{zhao_guided_2025, huang_adaptutar_2021}. Insights from this domain could be applied to the personal repair context. In an ongoing work, we are exploring the relation between a user's skill level and the modalities and levels of detail during a repair task.

\subsection{Testing}
In the final step, the repaired object has to be tested. Fabrication research can guide users through the process of ensuring their device or object is functional and identify potential issues to look out for. A challenge for this step, especially, is managing the very individual nature of repair: Every task is highly specific, and thus, tools need to be highly adaptable. 

\section{Conclusion} %mic drop 
As researchers, we want our work to matter, regardless of the field. In the context of fabrication research, we believe this can be achieved through finding and supporting meaningful use cases for personal fabrication.
The shift in public opinion on sustainable practices, along with the rise of the "right to repair" movement and supporting legislation show that repair is, in fact, a timely, relevant, impactful, and therefore meaningful use case for fabrication. 
While fabrication research has increasingly engaged with sustainability, repair is still often treated as a secondary outcome rather than a central design goal. As outlined in this paper, there is a wide range of opportunities and challenges in supporting novices, experts, and community repair spaces across the repair process. We therefore argue that future work should position repair as the main research agenda. In this emerging renaissance of repair, fabrication research has the opportunity to not merely respond but actively shape how repair is enabled, experienced, and valued.

%- Possible "vision": the toolbox of the future: no longer about treating tools and replacement parts separately: multi-purpose fabrication machine does it all

%- supporting repair shops/cafes
%- betting on the "predicted" resurgence of repair shops through new laws

%In the end, the argument is that we need to find meaningful use cases for personal fabrication and support people.
%- We dont need to develop products, thats not our job. our job is to generate knowledge

%\section{Authors and Affiliations}

%Each author must be defined separately for accurate metadata
%identification.  As an exception, multiple authors may share one
%affiliation. Authors' names should not be abbreviated; use full first
%names wherever possible. Include authors' e-mail addresses whenever
%possible.

%\begin{acks}
%To Robert, for the bagels and explaining CMYK and color spaces.
%\end{acks}

\bibliographystyle{ACM-Reference-Format}
\bibliography{bibliography}


\end{document}